\documentclass[superscriptaddress, aip, twocolumn]{revtex4}
\usepackage{color}
\usepackage{natbib}
\usepackage{graphicx}
\usepackage{amsmath}
\usepackage{ulem}
\begin{document}
\parskip 0pt plus 0pt

\title{Molecular dynamics of n-hexane: A quasi-elastic neutron scattering study on the bulk and spatially nanochannel-confined liquid}

\author{Tommy~Hofmann}
\email[]{tommy.hofmann@helmholtz-berlin.de}
\affiliation{Experimental Physics, Universit\"{a}t des Saarlandes, D-66041 Saarbr\"ucken,
  Germany}
\affiliation{Institute for Complex Magnetic Materials, Helmholtz-Zentrum Berlin, D-14109 Berlin, Germany}

\author{Dirk~Wallacher}
\affiliation{Experimental Physics, Universit\"{a}t des Saarlandes, D-66041 Saarbr\"ucken,
  Germany}
\affiliation{Department for Sample Environment, Helmholtz-Zentrum Berlin, D-14109 Berlin, Germany}

\author{Maria~Mayorova}  
\affiliation{J\"ulich Centre for Neutron Science, Forschungszentrum J\"{u}lich, D-52425 J\"{u}lich, Germany}

\author{Reiner~Zorn}
\affiliation{J\"ulich Centre for Neutron Science, Forschungszentrum J\"{u}lich, D-52425 J\"{u}lich, Germany}

\author{Bernhard~Frick}
\affiliation{Institut Laue-Langevin, BP 156, 38042 Grenoble Cedex 9, France}

\author{Patrick~Huber}
\email[]{patrick.huber@tuhh.de}
\affiliation{Experimental Physics, Universit\"{a}t des Saarlandes, D-66041 Saarbr\"ucken, Germany}
\affiliation{Pontificia Universidad  Cat\'{o}lica de Chile, Departamento de F\'{i}sica, Santiago, Chile}
\affiliation{Institute of Materials Physics and Technology, Hamburg University of Technology, D-21073 Hamburg, Germany}
\date{\today}

\begin{abstract}
We present incoherent quasi-elastic neutron scattering measurements in a wavevector transfer range from 0.4~$\AA^{-1}$ to 1.6$\AA^{-1}$ on liquid n-hexane confined in cylindrical, parallel-aligned nanochannels of 6~nm mean diameter and 260~$\mu$m length in monolithic, mesoporous silicon. They are complemented with, and compared to, measurements on the bulk system in a temperature range from 50~K to 250~K. The time-of-flight spectra of the bulk liquid can be modeled by microscopic translational as well as fast localized rotational, thermally-excited, stochastic motions of the molecules. In the nano-confined state of the liquid, which was prepared by vapor condensation, we find two molecular populations with distinct dynamics, a fraction which is immobile on the time scale of 1~ps to 100~ps  probed in our experiments and a second component with a self-diffusion dynamics slightly slower than observed for the bulk liquid. No hints of an anisotropy of the translational diffusion with regard to the orientation of the channels' long axes have been found. The immobile fraction amounts to about 5\% at 250~K, gradually increases upon cooling and exhibits an abrupt increase at 160~K (20~K below bulk crystallization), which indicates pore freezing.  
\end{abstract}
%\keywords{}
\maketitle
\section{Introduction}

The stochastic, thermally-excited motions within molecular liquids spatially confined on the nanometer scale is still attracting great interest both in fundamental and applied sciences.\cite{Zorn2010, McKenna_2010_a} In particular, self-diffusion of molecular liquids confined in mesoporous hosts plays a dominant role in catalysis and adsorptive separation.\cite{Schueth2002} It has also been discussed with regard to drug delivery applications \cite{Bras2011} and the origin of adsorption-desorption hysteresis found in mesoporous materials.\cite{Valiullin_2006_a} In terms of fundamental science the Brownian motion of molecules in pores a few nanometers across or in microchannels is interesting, since concepts concerning random motions in highly confined, crowded geometries \cite{Burada2009} or for molecules exposed to external fields (i.e. interaction potentials with the confining walls) can be scrutinized. \cite{Scheidler2002, Noeding2012,Bock2007, Sangthong2008, Krishna2009} Moreover, the direct relation between stochastic motions and viscous properties of a liquid (expressed by the Stokes-Einstein equation) allows one to explore the fluidity of liquids in nano-scale structures and in the proximity of the confining solid walls by studies of their self-diffusion dynamics. This is of obvious importance in the emerging field of nano- and microfluidics, where the exploration of these properties can be experimentally very demanding. \cite{Chan1985, Georges1993, Ruths1999, Becker2003, Eijkel2005, Neto2005, Joly2006, Huber2007, Kusmin2010, Walz2011, Bocquet2010, Gruener2011}\\
There is a sizeable number of experimental techniques, most prominently nuclear magnetic resonance, gravimetric uptake measurements, x-ray and light correlation spectroscopy as well as quasi-elastic neutron scattering, which allow one to study diffusion dynamics in nano- and mesoporous matrices. The advent of monolithic mesoporous hosts with tailorable channels a few nanometers across, e.g. mesoporous alumina and silicon membranes with parallel-aligned channels, have additionally increased the interest in this field and extended the analytical opportunities, for it allows a better and/or simpler comparison of theory and experiment. \cite{Gruener2008, Gruener2009, Kusmin2010}\\ 
Many studies of diffusion in restricted geometries are aimed at a comparison to the spatially unconfined, i.e. bulk, system. Over the years it has turned out, however, that this comparison is often hampered, if one has to solely rely on diffusion data reported in the literature and extracted with other measurement techniques. Even for identical bulk systems different experimental methods can deliver deviating dynamical properties and upon confinement these differences in the measured quantities can even be more pronounced. \cite{Mitra1992, Mitra1993, Chmelik2011} Different methods probe different length scales and thus also different time scales of diffusion processes. Translational self-diffusion in a crowded melt is, however, a highly cooperative phenomenon and can necessitate large scale molecular rearrangements depending on the investigated diffusion length. This translates to a dependency of the diffusion dynamics on the diffusion length, and thus diffusion time investigated. This was particularly convincingly demonstrated by seminal Molecular Dynamics simulations on unconfined, molecular liquids \cite{Alder1970, Alder1970a} and is all the more of importance for spatially nano-confined liquids, where the bare geometrical restriction can hamper both the movements of the diffusing molecule as well as the necessary mesoscopic rearrangement of the surrounding molecules. \cite{Cui2005, Devi2010}\\ 
This study tries to circumvent the challenge with respect to the comparability of self-diffusion properties gained by different methods by an investigation of both the bulk and spatially confined dynamics with an identical technique under similar experimental conditions. We report Incoherent Quasi-Elastic Neutron Scattering (IQENS) measurements on liquid n-hexane in tubular channels of 6~nm, that is nine times the length of the molecule. Hence, no single-file diffusion behavior is expected, however the interaction with the attractive confining walls is expected to affect the Brownian dynamics. \cite{Hansen1998, Kimmich1996, Stapf1995} Previous quasi-elastic neutron scattering measurements on liquid hexane in mesoporous silica were interpreted in terms of pronounced reductions in the translational self-diffusion of the molecules in the pore center (by an order of magnitude in the translational diffusion coefficient) and an even more slowed-down component in the pore wall vicinity. \cite{Baumert_2002_a} This study was performed with sponge-like, porous silica gel glasses and hence no investigation of anistropic diffusion with regard to the long pore axes was possible, an additional goal of this study.\\ 
Liquid hexane was a perfect candidate to study molecular motions in a bulk liquid (BL) and spatially confined liquid (CL) with IQENS techniques at moderate temperatures. It is a volatile room temperature liquid and in a neutron beam its molecules cause mainly incoherent scattering due to the high content of hydrogen atoms. Other scattering contributions from hydrogen or carbon atoms are not significant. Porous silicon was the preferred host to investigate the CL. It offered a natural environment to study the effect of nano-confinement and reduced dimensionality on stochastic motions. Its tubular/cylindrical pores have diameters of only a few nanometers but extend to lengths of a few hundred microns therefore creating an effective confinement in two dimensions for guest molecules like hexane. Its pores are parallel-aligned along the $<100>$-direction in single crystalline wafer which in combination with the huge length to diameter ratio not only raises the question whether molecular motions along the pore axis are different from radial motions but also offers the unique opportunity to study these potential anisotropies in selected scattering geometries as discussed in more detail below. 
\section{Sample Preparation}
Liquid hexane ($99.99\%$) was stored in sealed bottles in a dry atmosphere of a glove box. Bottles were only removed and opened to prepare BL and CL for the experiments. In this way contamination of hexane with water, which slowly builds up in a humid environment was kept at a minimum.
\\
Electrochemical anodization in hydrofluoric acid  was applied to etch porous structures in $500~\mu$m thick single crystalline silicon $<100>$-plates. Anodization parameters as wafer doping, electrolyte concentration, etching time and current were adjusted to facilitate the growth of $260~\mu$m thick porous layers whith a total porosity of $45\%$ at a average pore diameter of $6~nm$.
Etched substrates were stored in a container with distilled water to prevent contamination with hydrocarbons and only removed for experiments. 
\\
The BL as well as CL sample were prepared directly at the beamline. The setup to prepare BL included a traditional helium cryostate, a standard sample stick and a flat aluminum cell. The indium sealed aluminum cell with a width of 30~mm, a height of 50~mm and a depth of 0.5~mm contained next to the BL several sheets of thin aluminum to reduce the effective thickness of the hexane sample to avoid multiple scattering.  
\\
Setup and procedure to prepare the CL sample were more sophisticated in order to avoid the formation of excess bulk liquid outside the pores in the filling process. We employed a sample stick equipped with a stainless steal capillary, an air-tight network of small pipes, gas containers, pressure gages and valves, subsequently refered to as  ``gas handling'' and a high vacuum pumping station. 
The sample cell was tailored to match the dimensions of the 30~mm wide, 50~mm high and 0.5~mm thick porous matrix. The stainless steal capillary at the stick acted as narrow pipeline between the sample in the aluminum cell and the external gas handling. The gas handling itself served as reservoir and distribution system for hexane vapor. 
\\
The sample cell was evacuated at $373~K$ to $5\times10^{-7}$~mbar before the pores were stepwise filled with liquid hexane. This removed residual water from the inherently hydrophobic porous silicon host. For filling purposes, the sample was kept at  a temperature of $T_{\rm 0}=273$~K, the gas handling remained slightly warmer at room temperature close to $T=300$~K. Then volumetrically controlled  sorption steps allowed to successively physisorb well defined  amounts of liquid hexane inside the pore space up to complete filling with $N_{0}$ hexane molecules at the saturation vapor pressure of  $p=p_{\rm 0}(T_{\rm 0})=61$~mbar.
\begin{figure}[ht!]
\includegraphics[width=0.45\textwidth,angle=0]{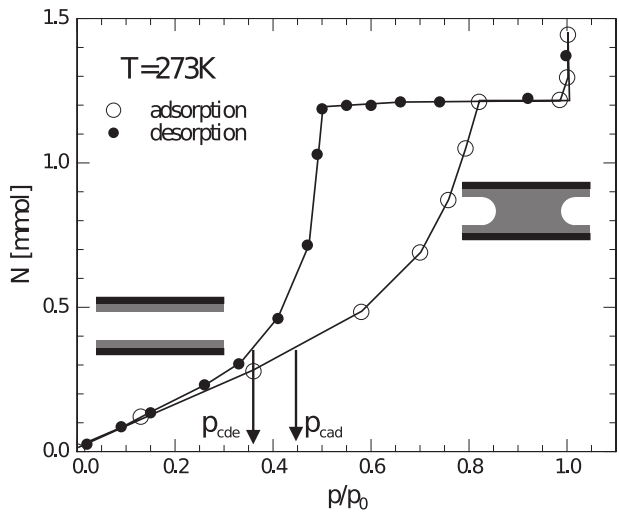}
\caption{ (Color online) Sorption isotherm for hexane condensed in porous silicon. The number of adsorbed molecules $N$ is shown as function of the reduced pressure $p/p_{\rm 0}$.
Open symbols refer to adsorption, closed symbols refer to desorption. Insets illustrate different geometric distributions of condensate in the pores as discussed in the text.}
\label{isotherm_plot}
\end{figure}
\\
A complete adsorption-desorption cycle for hexane in porous silicon is shown in Fig.~\ref{isotherm_plot}. It exhibits on ad- and desorption branch the number of physisorbed molecules $N$ as function of the reduced pressure $p/p_{\rm 0}$ and is commonly referred to as sorption isotherm. Two distinct sorption regimes on each branch are due to different geometric distributions of hexane molecules in the pores depending on the relative pressure. At low pressures, that is $p/p_{\rm 0}<p_{\rm cad}$ for adsorption and $p/p_{\rm 0}<p_{\rm cde}$ for desorption, 
the molecules physisorb preferentially close to the pore wall.\cite{Hair1969}  
In the hysteretic regime above the transition pressures  $p_{\rm cad}$ and $p_{\rm cde}$, which relate directly to the pore diameter, capillary condensation occurs in the pore centre \cite{Saam_1975_a}. A detailed discussion of the hysteresis ($p_{\rm cad}>p_{\rm cde}$) can be found in a multitude of references \cite{Saam_1975_a, Mason_1982_a, Mason_1983_a, Kornev_2002_a}.\\
For scattering experiments on CL's, the porous host was filled at $T_{\rm 0}=273$~K up to only $95\%$, decoupled from the gas handling and eventually brought on temperature $T<T_{\rm 0}$. Consequently all experiments were performed under quasi-isosteric conditions, and scattering contributions from coexisting BL could be ruled out definitely. If the pore volume was filled up to $95\%$ only at $273$~K and the cell was decoupled from the reservoir prior to cooling, the filling fraction of the host with hexane could not change by more than $2\%$ through condensation from coexisting vapor while cooling. The excess volume of cell and capillary, that is the combined volume of cell and capillary minus the volume of the empty sample,  were neglectable compared to the pore volume and did not contain enough hexane vapor to increase the filling of the host significantly. Conditions for bulk formation in the sample cell, namely  $p/p_{\rm 0}=1$ and $N/N_{\rm 0}>1$, were never fulfilled.\\
\section{Scattering Experiments}
Scattering experiments were performed at the time-of-flight spectrometer IN5 at Institut Laue-Langevin (Grenoble). The incoherent quasi elastic scattering of cold neutrons ($E_{\rm i}=2.08$~meV, $\lambda_{\rm i}=0.627$~nm) served as probe for stochastic motions of hexane molecules in the liquid phase. The scattering signal $S(Q,\omega)$, that is the number of neutrons scattered per second into a small solid angle $d\Omega$ with an energy transfer between $\hbar \omega$ and 
$\hbar(\omega+d\omega)$, was recorded as a function of wave vector transfer $Q$ and energy transfer $\hbar \omega$. Wave vector transfers $Q$ between $0.4$~\AA$^{\rm -1}$ and $1.6$~\AA$^{\rm -1}$ were accessible with a spatially fixed detector bank. The accessible energy range was $Q-$dependent. For the smallest $Q$ energy transfers $\hbar \omega$ were simultaneously discerned between $-1.5$~meV and  $+1.5$~meV with a high resolution of $\hbar\Delta\omega<\pm30~\mu$eV. For the largest $Q$ energy transfers $\hbar \omega$  were ascertained between  $-1.5$~meV and  $+8$~meV with a similar good resolution. Scattering data were taken at various temperatures $T$ to elucidate its effect on the dynamics.\\
For  BL scattering data were collected in a transmission geometry with the sample normal parallel to the incident neutron beam. Stochastic motions were studied at four temperatures above the bulk freezing point of $T_{\rm fb}=180$~K: $250$~K, $230$~K, $210$~K and $185$~K. At each temperature data were collected for at least $4$~h to guarantee good signal to noise ratios.\\
For CL two different scattering geometries were used to discern between radial and axial motions of hexane molecules in the cylindrical pores. In the first geometry the angle between pore axis (wafer normal) and incident beam was  $\Phi=135^{\circ}$. Therefore the main component of the probed wave vector transfers pointed along the symmetry axis of the pores, creating a high sensitivity to axial motions. In the second, complementary geometry the angle was $\Phi=45^{\circ}$. Probed wave vector transfers were nearly perpendicular to the pore axis. Therefore the scattering experiment elucidated mainly radial motions of hexane molecules in the pores. 
Scattering data were taken in the liquid regime above the bulk freezing temperature at $250$~K, $230$~K, $210$~K and $185$~K as well as below the bulk freezing point at $165$~K. Measurements of liquid hexane at $T=165$~K  were possible because of the sizeable shift of the liquid-solid phase transition upon spatial confinement,\cite{Knorr_2008_a} a shift which is for hexane not less than $20$~K \cite{Henschel_2009_a} in $6$~nm wide silicon pores. Each individual measurement took again $4$~h. The liquid-solid transition itself in confined hexane was probed by a sequence of 'quick-shot' measurements recorded while cooling the sample down from $165$~K to $10$~K. In steps of $1$~K data were sampled in each case for $5$~min.\\
Complementary scattering data were taken to allow and facilitate a quantitative data analysis. Background data were recorded for an empty aluminum cell and an evacuated aluminum cell containing the porous silicon plate. These data sets were used to adjust the experimental results for BL and CL with regard to scattering contributions not originating from the hexane pore condensate. Additional scattering data were taken from solid hexane at $T=10$~K. Here the complete incoherent  scattering signal builts up in an elastic line due to disrupted molecular motions at low temperatures and obtained data can be used for resolution corrections.
\section{\label{}Experimental Results}
\begin{figure}[t]
\includegraphics[width=0.33\textwidth,angle=0]{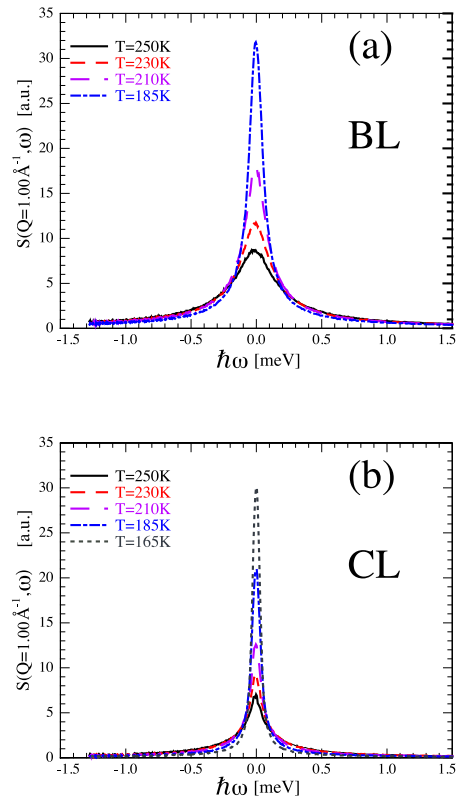}
\caption{(Color online) Incoherent inelastic scattering intensity $S(Q,\omega)$ as function of energy transfer $\hbar\omega$ and temperature $T$ for a fixed momentum transfer of $Q=1.00$~\AA$^{-1}$. Data are shown for BL (a) and for CL (b) with $\Phi=135^\circ$. }
\label{data_plot}
\end{figure}

This section is devoted to present a subset of the scattering data, which were recorded for BL's and CL's in the three-dimensional parameter space of wave vector transfer $Q$, energy transfer $\hbar\omega$ and temperature $T$. The presentation focuses for BL's and CL's on representative scattering data recorded for a wave vector transfer of $Q=1.00$~\AA$^{\rm -1}$. In the case of CL's the presentation is abridged and only scattering data sensitive to axial motions in the pores ($\Phi=135^{\circ}$) are discussed. Scattering data recorded for different momentum transfers or recorded for different scattering geometries ($\Phi=45^{\circ}$) could be discussed equivalently. They do not show qualitatively different characteristics as the ones discussed below.\\
Fig.~\ref{data_plot}a shows the incoherent quasi elastic scattering signal $S(Q,\omega)$ recorded for bulk hexane at various temperatures $T$ for a fixed momentum transfer of $Q=1.00$~\AA$^{\rm -1}$. Neutron intensities are shown as function of energy transfers $\hbar\omega$ between $-1.5$~meV and $+1.5$~meV.
Shown data were corrected for background scattering but not for energy resolution $R(Q,\omega)$. Neutrons which gain energy in the scattering process contribute to the intensity at $\hbar\omega>0$~meV, neutrons which lose energy contribute to the intensity at $\hbar\omega<0$~meV. Observed neutron intensities are always maximal for zero energy transfer ($\hbar\omega=0$~meV), that is elastic scattering. Intensities decay symetrically around the elastic line ($\hbar\omega=0$~meV). Positive and negative  half width at half maximum (HWHM) are identical apart from the sign as expected according to the principle of detailed balance for energy transfers $\hbar \omega << k_{\rm B}T$. Quantitatively the signals half width at half maximum decreases from $0.2$~meV at $T=250$~K to $0.07$~meV at $T=185$~K. This reflects qualitatively the expected mobility loss of the hexan molecules while approaching the freezing point.\\
Fig.~\ref{data_plot}b exhibits similar to Fig.~\ref{data_plot}a the incoherent quasi elastic scattering signal measured for hexane confined in the cavities of porous silicon. Data sets were again corrected for background contributions. A correction for energy resolution was not applied. The scattering signal exhibits the same qualitative behavior as discussed above for the BL. The signals half width at half maximum decreases from $0.09$~meV at $T=250$~K to $0.035$~meV at $T=165$~K indicating slowed dynamics close to the freezing point. Half widths for CL's  are at given temperature  about $50\%$ smaller than respective half widths for BL's. Molecular motions in the confined system are therefore in average slower than in the bulk system. A more detailed and quantitative interpretation of this important observation requires, however, a theoretical approach which accounts for different kinds of stochastic motions of hexane molecules in BL's and CL's  and must carefully consider the $Q-$dependent energy resolution $R(Q,\omega)$ of the experiment.
% ********************************************************************************
% ********************************************************************************
\section{\label{}Neutron Data Analysis: Theory}
Hexane molecules scatter
predominantly incoherently (due to the large incoherent cross-section of the protons). Therefore, we analyze in the following our data with a model for the incoherent scattering cross section $S_{\rm ic}(Q,\omega)$ which takes into account the different kinds of stochastic motions that may cause incoherent inelastic scattering in our experiments.\\
In the BL, long range translational diffusion as well as more localized motions, that are rotational reorientations of entire molecules or fast intra-molecular motions cause contributions to the inelastic scattering signal. Incoherent inelastic scattering due to translational diffusion is commonly described by a single Lorentzian term with $Q$-dependent HWHM $\gamma_{\rm T}(Q)$.\cite{Furrer_2009_a} It is:
\begin{equation}
S_{\rm ic}^{trans}(Q, \omega) = \frac{1}{\pi}\frac{\gamma_{\rm T}(Q)}{\hbar^2\omega^2+\gamma_{\rm T}(Q)^2}.
\end{equation}
For continuum diffusion as governed by Fick's second law, it can be derived that $\gamma_{\rm T}=\hbar D Q^{\rm 2}$ ,where $D$ refers to the so called self-diffusion constant. \cite{Furrer_2009_a} A more sophisticated, microscopic approach based on a jump diffusion model confirms the linear relationship between $\gamma_{\rm T}$ and $Q^{\rm 2}$ for small $Q$-values, but predicts deviations for larger $Q$'s. \cite{Furrer_2009_a, Chudley_1961_a}\\
Localized molecular reorientations and intra-molecular motions can approximately be governed by \cite{Smuda_2008_a, Bee_1992_a} 
\begin{equation}
S_{\rm ic}^{loc}(Q, \omega) = A_{\rm 0}(Q)\delta(\omega)+(1-A_{\rm 0}(Q))\frac{1}{\pi}\frac{\gamma_{\rm loc}}{\hbar^2\omega^2+\gamma_{\rm loc}^2}.
\end{equation}
$A_{\rm 0}(Q)$ is the so called elastic incoherent structure factor and emerges from spatial restrictions by definition imposed on localized stochastic motions. \cite{Bee_1992_a} The HWHM $\gamma_{\rm loc}$ of the Lorentzian term in Eqn.~(5) might be related to a characteristic relaxation time $\tau$ of localized motions by $\gamma_{\rm loc}=\hbar /\tau$ with  $\tau=\tau_{\rm 0}e^{E_{\rm A}/k_{\rm B}T}$. $E_{\rm A}$ is the activation energy of the considered processes. Assuming an independent superposition of long range diffusional processes and localized motions leads to the total incoherent scattering cross section:
\begin{align}
 \label{bulk_equation}
S_{\rm ic}^{total} (Q, \omega) = {} & S_{\rm ic}^{\rm trans}(Q, \omega) \otimes S_{\rm ic}^{\rm loc} (Q, \omega)\\
= {} & C*(\frac{A(Q)}{\pi}\frac{\gamma_{\rm T}}{\hbar^2\omega^2+\gamma_{\rm T}^2}\nonumber\\
- {} & \frac{1-A(Q)}{\pi}\frac{\gamma_{\rm T}+\gamma_{\rm loc}}
{\hbar^2\omega^2+(\gamma_{\rm T}+\gamma_{\rm loc})^2})\nonumber.
\end{align}
Here $\otimes$ denotes a convolution in $\omega$ and $C$ is a $Q$-independent scaling factor.\\
In the CL the stochastic motions will be not qualitatively different, but at least two different molecular surroundings have to be discerned. Molecules close to the pore walls are expected to exhibit different dynamic properties than molecules in the pore centre.\cite{Scheidler2002} The most obvious approach would therefore be to compose a total scattering function as sum of two terms, each similar to Eqn.~\eqref{bulk_equation}, which represent wall layers respectively capillary condensate. Such an approach, however, would inevitably lead to a multitude of free and strongly correlated parameters in the model function. Consequently we propose a simplified approach to reduce the number of parameters:  
\begin{align}
\label{confined_equation}
S_{\rm ic}^{\rm total}(Q,\omega) = {} & C*(
(1-f_{\rm c}) ( \frac{A(Q)}{\pi}\frac{\gamma_{\rm T}}{\hbar^2\omega^2+\gamma_{\rm T}^2}\\
- {} &  \frac{1-A(Q)}{\pi}\frac{\gamma_{\rm T}+\gamma_{\rm loc}}
{\hbar^2\omega^2+(\gamma_{\rm T}+\gamma_{\rm loc})^2} ) +\frac{f_{\rm c}}{\hbar}\delta(\omega))\nonumber
\end{align}
Here it is assumed that a fraction $1-f_{\rm c}$ of the CL can be described similar to a BL as sum of two Lorentzians, while the complementary fraction $f_{\rm c}$ of the CL is completely immobile and is therefore represented by a delta function $\delta(\omega)$ in the scattering cross section.\\
A convolution in $\omega-$space of  Eqn.~\eqref{bulk_equation} respectively Eqn.~\eqref{confined_equation} with the experimental resolution function $R(Q,\omega)$ allows to relate  theoretical models 
\begin{equation}
I(Q,\omega)=R(Q,\omega) \otimes S_{\rm ic}^{\rm total}(Q,\omega) 
\label{convolution_equation}
\end{equation}
and observed scattering signals for BL and CL.
% **********************************************************************************************
% **********************************************************************************************
\section{\label{}Neutron Data Analysis: Results}  
\begin{figure}[ht]
\includegraphics[width=0.33\textwidth,angle=0]{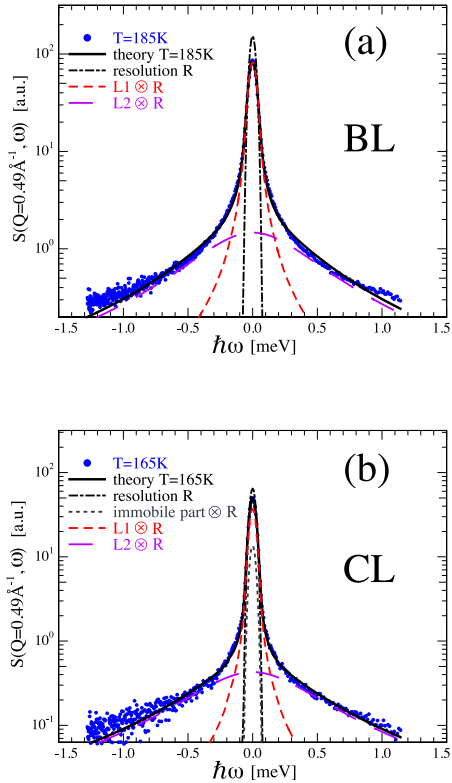}
\caption{(Color online) (a)~Symbols (circles) exhibit the incoherent inelastic scattering from BL at $T=185$~K. The solid line shows the result of  nonlinear square fitting utilizing Eqn.~\eqref{bulk_equation} and Eqn.~\eqref{convolution_equation}. The  short dashed line represents the first Lorentzian term in Eqn.~\eqref{bulk_equation} convoluted with the energy resolution function denoted as dot-dashed line. The long dashed line represents the second Lorentzian term in Eqn.~\eqref{bulk_equation} convoluted with the energy resolution function. (b)~Symbols (circles) exhibit the incoherent inelastic scattering signal from CL at $T=165$~K ($\Phi=135^{\circ}$). First and second Lorentzian  term of Eqn.~\eqref{confined_equation} were convoluted with the energy resolution function and are shown as short and long dashed lines. The dotted
line illustrates the $\delta$-function in Eqn.~\eqref{confined_equation} convoluted with the energy resolution. The solid line is the result of the fitting approach based on  Eqn.~\eqref{confined_equation} and Eqn.~\eqref{convolution_equation}.}
\label{fitting_plot}
\end{figure}

\begin{figure}[ht]
\includegraphics[width=0.4\textwidth,angle=0]{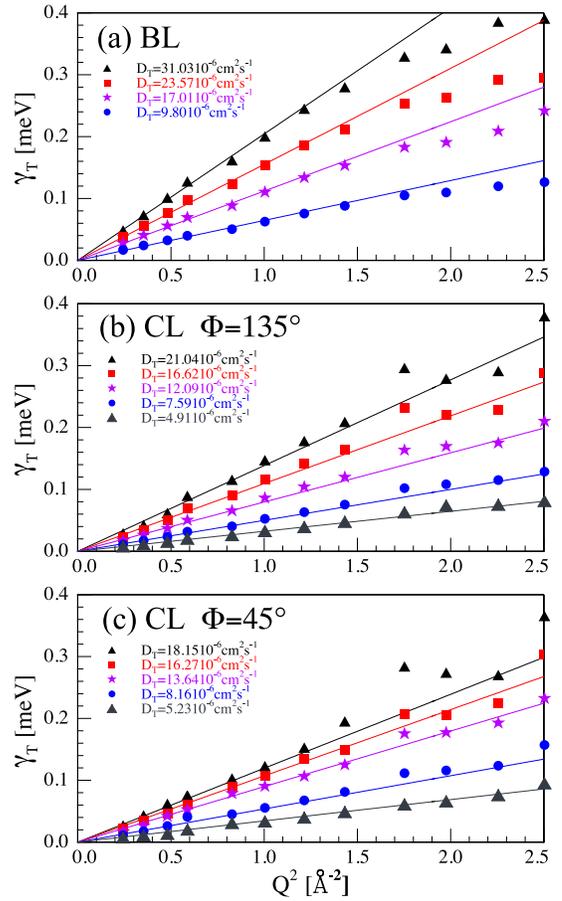}
\caption{(Color online) Half width at half maximum $\gamma_{\rm T}$ of the first Lorentzian term in Eqn.~\eqref{bulk_equation} respectively Eqn.~\eqref{confined_equation} as function of $Q^2$ and $T$. Data  are shown for isotropic BL (a), for CL with $\Phi=135^\circ$ (b) and for CL with $\Phi=45^\circ$ (c). Different  symbols (black triangle, square, star, circle, grey triangle) refer to different temperatures (250~K, 230~K, 210~K, 185~K, 165~K). Solid lines are linear fits to estimate the translational diffusion constant $D_{\rm T}$. The fitting range was restricted to $0\leq Q^2/$\AA$^{-2}\leq1$.}
\label{HWHM_q2_plot}
\end{figure}

\begin{figure}[ht]
\includegraphics[width=0.35\textwidth,angle=0]{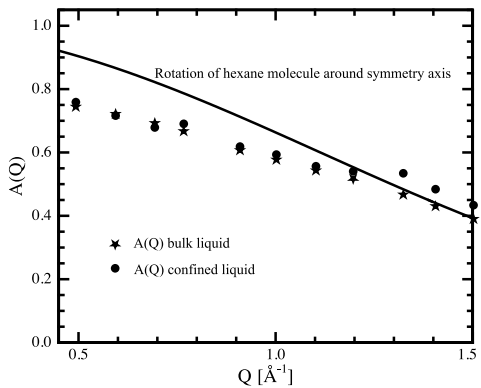}
\caption{(Color online) Elastic incoherent structure factor as  measured for bulk hexane (stars) and confined hexane (circles). Expected elastic incoherent structure factor (solid line) for rotation of hexane molecules around their symmetry axis.}
\label{EISF_bulk_plot}
\end{figure}

Nonlinear least square fitting based on the discussed models was utilized to analyze the scattering data quantitatively in the probed $T-$ and $Q-$range.  Eqn.~\eqref{bulk_equation} and Eqn.~\eqref{convolution_equation} were adjusted to the BL data. In the fitting approach the HWHM $\gamma_{\rm T}$ of the first Lorentzian in Eqn.~\eqref{bulk_equation}, the characteristic relaxation time $\tau$ for fast and localized motions, the elastic incoherent structure factor $A(Q)$  as well as the scaling factor $C$ were treated as free parameters. \\
Fig.~\ref{fitting_plot}a  illustrates representatively the success of the applied fitting approach for a particular set of $Q$ and $T$. In the semi-logarithmic plot symbols exhibit  the incoherent scattering signal $S(Q,\omega)$ taken at $Q=0.49$~\AA$^{-1}$ and $T=185$~K corrected for background contributions. The solid line shows the best approximation of our model (Eqn.'s~\eqref{bulk_equation}-\eqref{convolution_equation}) to the data. Different contributions to the model function \eqref{bulk_equation} are shown as dashed lines. The Lorentzian term with HWHM $\gamma_{\rm T}$ convoluted with the resolution function is shown as short-dashed line. For small energy transfers ($\omega<0.2$~meV) this term is the main contribution to the scattering signal. 
The long-dashed line represents the second Lorentzian  with HWHM $\gamma_{\rm T}+\gamma_{\rm loc}$ convoluted with the resolution function. It is the main contribution to the scattering signal for large energy transfers $\omega>0.5$~meV. 
The resolution function itself is shown as a dot-dashed line. 
Fitting of scattering data at other points in the parameter space $Q$ and $T$ gives similar good results. A detailed discussion of the obtained fitting parameter and their dependence on $Q$ and $T$ will be given in the subsequent paragraphs.\\  
In the fitting approach the parameter $C$ does not show a systematic dependence on the wave vector transfer $Q$. Independent of temperature $T$  found values for $C$ are close to unity. Deviations from unity are smaller than $10\%$ and as a direct consequence of Eqn.~\eqref{bulk_equation} and Eqn.~\eqref{convolution_equation} the in $\omega-$space integrated intensity of the scattering signal at given $Q$ is considered to be equal to the integrated intensity of the respective energy resolution function $R(Q,\omega)$. The latter is in so far not surprising as we used the scattering signal from the solidified BL as resolution function  in the  fitting approach and not the scattering signal from standard vanadium samples. Constance of $C$  illustrates only the conservation of incoherent scattering intensity at a particular $Q$. It is therefore a first successful consistency check for our model. 
\\
Fig.~\ref{HWHM_q2_plot}a shows the HWHM $\gamma_{\rm T}$ of the first Lorentzian in Eqn.~\eqref{bulk_equation} as function
of  the square of the wave vector transfer $Q$ for probed temperatures between $250$~K and $185$~K. For small wave vector transfers $0<Q/$\AA$^{-1}<1$ the HWHM $\gamma_{\rm T}$ is at given temperature $T$ proportional to $Q^2$ as expected for translational continuum diffusion.\cite{Furrer_2009_a} Linear fitting in this restricted $Q-$range was exploited to extract self diffusion coefficients $D_{\rm T}$  according to  $\gamma_{\rm T}=\hbar DQ^2$. Obtained temperature dependent diffusion coefficients $D_{\rm T}$ are listed in the Fig.~\ref{HWHM_q2_plot}a and Table \ref{diffusion_constants}. The diffusion coefficients for BL vary between $1\times10^{-5}$~cm$^2$s$^{-1}$ at $185$~K and $3.1\times10^{-5}$~cm$^2$s$^{-1}$ at $250$~K. 
Diffusion coefficients between  $0.5\times10^{-5}$~cm$^2$s$^{-1}$ at $185$~K and $2.15\times10^{-5}$~cm$^2$s$^{-1} $ at $250$~K have been found for bulk hexane in NMR spin-echo measurements performed by Douglass and McCall 1958 and 1959.\cite{Douglass_1958_a, McCall_1959_a}
\\
Deviations form the linear relationship  between $\gamma_{\rm T}$ and $Q^2$ at larger  scattering vectors $Q$ are expected. They mark  length scales $2\pi/Q$ on which molecular diffusion can not be described any more by macroscopic models (Fick's law) but more microscopic approaches \cite{Chudley_1961_a} are required.
\\
% ***********************************************************************************
% ********************************************************************************************************
\begin{table}[htdp!]
\caption{Translational diffusion constants for isotropic bulk hexane ($D_{\rm BL}$) and for confined hexane along, respectively perpendicular to the pore axis ($D_{\rm CL}^{135}$, $D_{\rm CL}^{45}$) measured at different temperatures $T$.}
\begin{center}
\begin{tabular}{| c | c | c | c | c |}
\hline                       
  T [K] & $D_{\rm BL}$ [cm$^{2}$s$^{-1}$] & $D_{\rm CL}^{135}$  [cm$^{2}$s$^{-1}$] & $D_{\rm CL}^{45}$  [cm$^{2}$s$^{-1}$] & $D_{\rm CL}^{135}/D_{\rm BL}$ \\ \hline \hline
  250 & $3.1\times10^{-5}$ & $2.1\times10^{-5}$ & $1.8\times10^{-5}$ & 0.68 \\ \hline 
  230 & $2.4\times10^{-5}$ & $1.7\times10^{-5}$ & $1.6\times10^{-5}$ &  0.71 \\ \hline
  210 & $1.7\times10^{-5}$ & $1.2\times10^{-5}$ & $1.4\times10^{-5}$ & 0.71 \\ \hline
  185 & $1.0\times10^{-5}$ & $0.8\times10^{-5}$ & $0.8\times10^{-5}$ & 0.8  \\ \hline
  165 & N.A. & $0.5\times10^{-5}$ & $0.5\times10^{-5}$ & N.A. \\ 
  \hline  
\end{tabular}
\end{center}
\label{diffusion_constants}
\end{table}
% ***************************************************
In the fitting routine the paramater $\tau$ was considered independent of $Q$ as motivated by Dianoux et al..\cite{Dianoux_1975_a}
Obtained $\tau$ values suggest a time constant $\tau_0=0.4$~ps in the pico-second range and very low activation energies of  less than $10$~meV for the localized motions governed  by the second Lorentzian in the fitting function.
Therefore discussed, probably intra-molecular motions appear to be one to two orders in magnitude faster than translational diffusion with $T-$dependent relaxation times $\tau_{\rm T}=<l^2>/6D$ between  $20$~ps and $70$~ps.
Here $<l^2>$ denotes the mean square displacement of a hexane molecule in the time $\tau_{\rm T}$ and was assumed to be given by the square of the nearest neighbor distance ($d_{\rm NN}=0.65$~nm) in the liquid.
\\
The data analysis did not reveal any significant temperature dependence of the elastic incoherent structure factor $A(Q)$ at a fixed wave vector transfer $Q$. The obtained $Q-$dependence is shown in Fig.~\ref{EISF_bulk_plot} (stars).  $A(Q)$ decreases continuously  from $0.75$ at $Q=0.5$~\AA$^{-1}$ to $0.4$ at $Q=1.5$~\AA$^{-1}$. 
The solid line exhibits an expected decay in $A(Q)$ for rotational diffusion of hexane molecules around their symmetry axis.\cite{Dianoux_1975_a} Although it shows the same qualitative behavior as the data a quantitative agreement is not achieved. This is in so far not surprising as a multitude of different localized motions in the hexane molecule define $A(Q)$ not only rotations around the symmetry axis. Further multiple scattering effects as outlined by Zorn et al. \cite{Zorn_2002_a} might lead to this discrepancy between experiment and theory.
\\
Eqn.~\eqref{confined_equation} and Eqn.~\eqref{convolution_equation} facilitated a quantitative  analysis of  the CL data. Free parameter in the fitting routine were the HWHM $\gamma_{\rm T}$, the HWHM $\gamma_{\rm loc}$, the Elastic Incoherent Scattering Factor $A(Q)$, the scaling parameter $C$ and the fraction $f_{\rm c}$ of immobile hexane molecules in the pores.
\\
Fig.~\ref{fitting_plot}b exhibits the excellent agreement found between CL data ($\Phi=135^{\circ}$) and model function (Eqn.~\eqref{confined_equation}) for a selected combination of wave vector transfer $Q=0.49$~\AA$^{-1}$ and temperature $T=165$~K.  Scattering signal $S(Q, \omega)$ and best approximation based on our model are shown, as well as the different contributions to the model function that were discussed above.  As for bulk hexane translational diffusion and fast localized motions contribute significantly to the model function. But in order to facilitate a good approximation of the scattering signal at small energy transfers $\hbar\omega$ a percentage $f_{\rm c}>0$ of hexane molecules is set in the fitting routine to be immobile.  Scattering from this immobile molecules is  shown as dotted line in the figure. It is a vital contribution to the model function in an energy range  around the elastic line ($\hbar\omega=0$~meV) which corresponds to the half width half maximum of the energy resolution function $R(Q, \omega)$ at given $Q$.
Applying our fitting approach  to data that were recorded at other points in the parameter space $Q,T$, respectively for the second probed scattering geometry ($\Phi=45^{\circ}$), leads to similar compelling results.~A quantitative and systematic evaluation of the free parameters obtained by fitting the data will be given in the subsequent paragraphs.
\\
In the fitting routine parameter $C$ exhibits the same quantitative behavior as already observed in the analysis of the BL data.  Independent of wave vector transfer $Q$ and  temperature $T$  values for $C$ were found which do not deviate by more than $10\%$ from unity.  
The elastic incoherent structure factor $A(Q)$ found for the CL does not show any significant temperature dependence. Its $Q-$dependence in Fig.~\ref{EISF_bulk_plot} (circles) exhibits exactly  the same characteristics as observed for the BL.  
Paramter $\tau$ predicts again relaxations times in the order of 1 ps and activation energies below 10 meV for the fast localized motions.
\\
 Fig.~\ref{HWHM_q2_plot}b and Fig.~\ref{HWHM_q2_plot}c exhibit  the relation ship between $\gamma_{\rm T}$ and the square of the wave vector transfer $Q$ for various temperatures $T$. Frame b relates to scattering data sensitive to motions along the pore axis ($\Phi=135^{\circ}$). Frame c relates to scattering data sensitive to radial motions in the pore ($\Phi=45^{\circ}$). The linearity between $\gamma_{\rm T}$ and $Q^2$ in the wave vector range  $0<Q/$\AA$^{-1}<1$ was exploited to extract translational self diffusion constants $D_{\rm T}$.~Resulting diffusion coefficients for the different scattering geometries are listed in the respective figures as well as in Table~\ref{diffusion_constants}. For $\Phi=135^{\circ}$  diffusion coefficients from $2.1\times10^{-5}$~cm$^2$s$^{-1}$ to $0.5\times10^{-5}$~cm$^2$s$^{-1}$ are found in the temperature range between $250$~K and $165$~K. For $\Phi=45^{\circ}$ diffusion constants between $1.8\times10^{-5}$~cm$^2$s$^{-1}$ and $0.5\times10^{-5}$~cm$^2$s$^{-1}$ are found. 
\\
Self diffusion coefficients  in the mobile part of the pore condensate are about $20\%$ to $30\%$ smaller than respective ones in liquid bulk hexane. In the margin of error a directional anisotropy can not be ascertained. Translational diffusion along the pore axis is not quantitatively different from translational diffusion perpendicular to the pore axis.
Radial confinement on a length scale of 6~nm (diameter of the pore), that is 9 times the length of an hexane molecule in an all-trans configuration, seems  not to be small enough compared to the approximated displacement $\sqrt{<l^2>}\approx 0.65$nm of a single hexane molecules in the time $\tau_{\rm T}$ to affect radial diffusion coefficients .  
\\
% *************************************************************************************************
% *************************************************************************************************
% ********************************************************************************************************
\begin{figure}[ht]
\includegraphics[width=0.45\textwidth,angle=0]{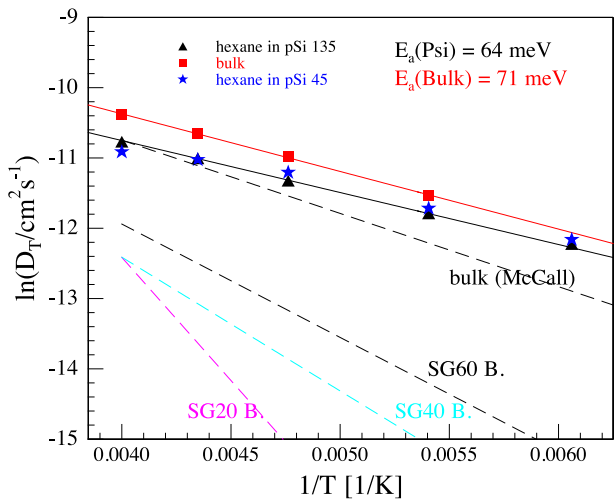}
\caption{(Color online) Arrhenius plot for the $T$-dependence of the translational diffusion constant $D_{\rm T}$: Squares denote bulk data. Triangles and stars refer to  diffusion constants along the pore axis ($\Phi=135^\circ$) respectively perpendicular to the pore axis ($\Phi=45^\circ$). Solid lines  illustrate linear fits to extract the activation energies $E_{\rm A}$. Dashed lines exhibit IQENS results from Baumert et al. as shown in \cite{Baumert_2002_a} for mesoporous silica gels with $2$~nm, $4$~nm and $6$~nm pore diameter and results form McCall et al. obtained in NMR spin-echo experiments.\cite{McCall_1959_a} Results of Baumert et al. are not corrected for a missing factor of $2\pi$.}
\label{arrhenius_plot}
\end{figure}
% ********************************************************************************************************
\begin{figure}[ht]
\includegraphics[width=0.5\textwidth,angle=0]{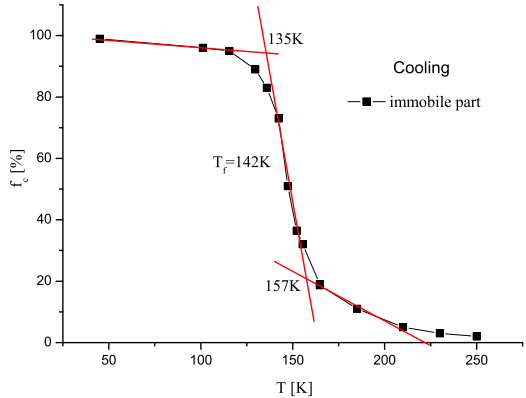}
\caption{(Color online) Temperature dependence of the amount $f_{\rm c}$ of immobile hexane in the pores. Symbols represent values extracted from the scattering data. Lines help to guide the eye.}
\label{immobile_part_plot}
\end{figure}
% **********************************************************************************************************
The Arrhenius plot in Fig.~\ref{arrhenius_plot} exhibits self diffusion constants $D_{\rm T}$ reported in this and various other articles for bulk hexane and confined hexane as function of inverse temperature $1/T$. Symbols illustrate the diffusion constants reported in this article for BL and the mobile part of the CL. 
Activation energies as estimated from the arrhenius plot are similar for BL and CL. The obtained value of $60$~meV$-70$~meV is $20\%$ smaller than the one reported by McCall.\cite{McCall_1959_a}  Fig.~\ref{arrhenius_plot} shows also Diffusion constants reported by Baumert et al. \cite{Baumert_2002_a} for hexane confined in mesoporous silica gels with pore diameters of $2$~nm, $4$~nm and $6$~nm. The reported constants are significantly different from our results. However, this difference can be traced back to a missing factor of $2\pi$ in the calculations in the before mentioned article. The reported activation energies for tranlational diffusion in \cite{Baumert_2002_a} are considerably larger than the one found here. This trend is partially explained by the pore-size dependence of the activation energy as illustrated in.\cite{Baumert_2002_a} 
% **********************************************************************************************************
\\
An analysis of scattering data in the temperature range between $250$~K and $40$~K revealed a temperature dependent fraction $f_{\rm c}$ of hexane in the pores (Fig.~\ref{immobile_part_plot}), which is immobile on the probed time scale of 1 ps to 100ps. At $250$~K about $5\%$ of the CL showed no dynamics. This is significantly less material as required to entirely cover   cylindrical pore walls with radius $R=3~nm$. Therefore  it seems likely that this fraction of hexane represents molecules that are trapped in deep adsorption sites of the less then perfectly smooth, but rough internal pore walls.
With decreasing temperature $f_{\rm c}$ increased at first slowly. Close to $T=160$~K, that is the freezing point of hexane in $6$nm silicon pores as reported in \cite{Henschel_2009_a} about $20\%$ of the CL appeared to be immobile. 
This agrees fairly well with the amount of hexane on the pore walls at onset of capillary condensation and makes a strong case that hexane molecules in close vicinity of the walls lose their mobility while the CL approaches the freezing point. This population corresponds likely to a flat-lying, strongly adsorbed monolayer of hexane molecules, which sticks to the pore wall, similarly as it has been inferred from ellipsometric and rheological measurements for medium-length n-alkanes in the proximity of silicon and silica surfaces. \cite{Chan1985, Volkmann2002, Gruener2009, Kusmin2010}
Here it should however be noted that between $180$~K and $160$~K hexane already solidifies in a small number of pores with a larger than average diameter ($>6$~nm). These immobile hexane molecules will also contribute to $f_{\rm c}$ as the scattering experiment can not discern between hexane stuck on the pore walls and hexane solidified in larger than average pores.   
Below $160$~K the immobile fraction increases rapidly. The hexane solidifies in the pores. This solidification process spreads over an temperature range of $20$~K associated with the pore size distribution. Below $130$~K there is no mobile hexane left. The fraction $f_{\rm c}$ of immobile hexane was extracted from data which were recorded while the sample was cooled down.\\
\section{\label{}Conclusions}
We performed time-of-flight experiments at spectrometer IN5 (Grenoble) to probe stochastic motions in liquid bulk hexane and liquid hexane confined in porous silicon which exhibits almost cylindrical pores parallel-aligned along the surface normal with an average pore diameter of $6$~nm. \\
Long range as well as localized motions of hexane molecules confined in the porous silicon matrix were found to be slightly slower than in the BL. A temperature dependent fraction $f_{\rm c}$ of the CL was found to be immobile on the probed time scales. At $250$~K about $5\%$ of the confined hexane, did not show any dynamics at all. These $5\%$ of hexane likely accounts for molecules pinned in deep adsorption sites of the rough pore walls. At $165$~K before onset of freezing about $20\%$ of the pore filling, that is the amount of hexane on the pore walls before onset of capillary condensation, was immobile.\\ 
In the mobile part ($1-f_{\rm c}$) of the CL translational diffusion constants $D_{\rm T}$ were reduced by $20\%$ to $30\%$  compared to the BL, but not  by an order of  a magnitude as inadvertently claimed by Baumert et al..\cite{Baumert_2002_a} A reason for the slowed dynamics in the mobile hexane can not entirely be elucidated by the performed experiments and analysis. It might be a direct result of  spatial confinement imposed on  molecules in the pores or to some extent an artifact of the simplified fitting approach which only discerns  immobile hexane molecules and mobile hexane molecules. A radial gradient in the diffusion constants, that is bulk diffusivity in the pore centre and a decreasing diffusivity closer to the pore walls, seems to be intuitive and suggested both by Molecular Dynamics simulations and experiments.\cite{Scheidler2002,Schranz2007} But attempts to apply models that account more clearly for a radial dependence of dynamics failed due to a multitude of strongly correlated fitting parameters and were therefore not suited to perform a reliable and meaningful analysis of the scattering data.\\
Follow-up experiments should exploit partial fillings of porous hosts with hexane to probe radial variations in diffusivity directly. Previous such studies employing mesoporous MCM-41 silica matrices indicated, however, that additional to the stochastic motions of the molecules in the pore centre new diffusion mechanisms, i.e. at the solid/vapor and at the liquid/vapor interfaces of the partially filled matrices, have to be considered.\cite{Hansen1998, Courivaud2000} They may hamper an extraction of the radial dependency by such filling-fraction-dependent measurements. Nevertheless, probing CL's before and after the onset of capillary condensation in the pores should allow to discern more clearly the dynamics of hexane molecules close to the pore walls and dynamics of hexane molecules close to the pore centre.\\
An anisotropy in diffusion coefficients was not found in the CL. Translational diffusion of hexane molecules along the pore axis was not different from diffusion perpendicular to the pore axis. A radial confinement on $6$nm was not sufficient to affect the diffusion  of hexane molecules which is characterized by a mean displacement of about $l=0.65$~nm on a time scale of $\tau_{\rm T}$. This finding is corroborated by high-resolution optical birefringence measurements on hexane and other rod-like molecules of comparable size confined in cylindrical nanochannels, which gave no hints of any static orientational order.\cite{Kityk2009, Wolff2010} Experiments on confined liquids in porous hosts with less wide pores are expected to show a different outcome. If the diameter of the pores gets closer to the average displacement $l$ of hexane molecules in the time $\tau_{\rm T}$ the diffusion in a cylindrical pore is expected to become more and more one dimensional and preferentially directed along the pore axis.\cite{Alder1970, Alder1970a, Cui2005, Bock2007, Kusmin2010, Kusmin2010a}\\
% **************************
\section{\label{}Acknowledgments}
This work was supported by the German Research Foundation (DFG) within the priority program 1144, Micro- and Nanofluidics, Grant No. Hu 850/2.


\begin{thebibliography}{57}
\expandafter\ifx\csname natexlab\endcsname\relax\def\natexlab#1{#1}\fi
\expandafter\ifx\csname bibnamefont\endcsname\relax
  \def\bibnamefont#1{#1}\fi
\expandafter\ifx\csname bibfnamefont\endcsname\relax
  \def\bibfnamefont#1{#1}\fi
\expandafter\ifx\csname citenamefont\endcsname\relax
  \def\citenamefont#1{#1}\fi
\expandafter\ifx\csname url\endcsname\relax
  \def\url#1{\texttt{#1}}\fi
\expandafter\ifx\csname urlprefix\endcsname\relax\def\urlprefix{URL }\fi
\providecommand{\bibinfo}[2]{#2}
\providecommand{\eprint}[2][]{\url{#2}}

\bibitem[{\citenamefont{Zorn et~al.}(2010)\citenamefont{Zorn, van Eijck, Koza,
  and Frick}}]{Zorn2010}
\bibinfo{author}{\bibfnamefont{R.}~\bibnamefont{Zorn}},
  \bibinfo{author}{\bibfnamefont{L.}~\bibnamefont{van Eijck}},
  \bibinfo{author}{\bibfnamefont{M.~M.} \bibnamefont{Koza}}, \bibnamefont{and}
  \bibinfo{author}{\bibfnamefont{B.}~\bibnamefont{Frick}},
  \bibinfo{journal}{European Physical Journal-special Topics}
  \textbf{\bibinfo{volume}{189}}, \bibinfo{pages}{1} (\bibinfo{year}{2010}).

\bibitem[{\citenamefont{McKenna}(2010)}]{McKenna_2010_a}
\bibinfo{author}{\bibfnamefont{G.~B.} \bibnamefont{McKenna}},
  \bibinfo{journal}{European Physical Journal-special Topics}
  \textbf{\bibinfo{volume}{189}}, \bibinfo{pages}{285} (\bibinfo{year}{2010}).

\bibitem[{\citenamefont{Sch\"uth and Sing}(2002)}]{Schueth2002}
\bibinfo{author}{\bibfnamefont{F.}~\bibnamefont{Sch\"uth}} \bibnamefont{and}
  \bibinfo{author}{\bibfnamefont{K.}~\bibnamefont{Sing}},
  \emph{\bibinfo{title}{Handbook of Porous Solids}}
  (\bibinfo{publisher}{Wiley-VCH}, \bibinfo{year}{2002}).

\bibitem[{\citenamefont{Bras et~al.}(2011)\citenamefont{Bras, Merino, Neves,
  Fonseca, Dionisio, Schoenhals, and Correia}}]{Bras2011}
\bibinfo{author}{\bibfnamefont{A.~R.} \bibnamefont{Bras}},
  \bibinfo{author}{\bibfnamefont{E.~G.} \bibnamefont{Merino}},
  \bibinfo{author}{\bibfnamefont{P.~D.} \bibnamefont{Neves}},
  \bibinfo{author}{\bibfnamefont{I.~M.} \bibnamefont{Fonseca}},
  \bibinfo{author}{\bibfnamefont{M.}~\bibnamefont{Dionisio}},
  \bibinfo{author}{\bibfnamefont{A.}~\bibnamefont{Schoenhals}},
  \bibnamefont{and} \bibinfo{author}{\bibfnamefont{N.~T.}
  \bibnamefont{Correia}}, \bibinfo{journal}{Journal of Physical Chemistry C}
  \textbf{\bibinfo{volume}{115}}, \bibinfo{pages}{4616} (\bibinfo{year}{2011}).

\bibitem[{\citenamefont{Valiullin et~al.}(2006)\citenamefont{Valiullin, Naumov,
  Galvosas, K\"arger, Woo, and Monson}}]{Valiullin_2006_a}
\bibinfo{author}{\bibfnamefont{R.}~\bibnamefont{Valiullin}},
  \bibinfo{author}{\bibfnamefont{S.}~\bibnamefont{Naumov}},
  \bibinfo{author}{\bibfnamefont{P.}~\bibnamefont{Galvosas}},
  \bibinfo{author}{\bibfnamefont{J.}~\bibnamefont{K\"arger}},
  \bibinfo{author}{\bibfnamefont{F.}~\bibnamefont{Woo}, \bibfnamefont{H.~J.
  andf~Porcheron}}, \bibnamefont{and} \bibinfo{author}{\bibfnamefont{P.~A.}
  \bibnamefont{Monson}}, \bibinfo{journal}{Nature}
  \textbf{\bibinfo{volume}{443}}, \bibinfo{pages}{965} (\bibinfo{year}{2006}).

\bibitem[{\citenamefont{Burada et~al.}(2009)\citenamefont{Burada, Haenggi,
  Marchesoni, Schmid, and Talkner}}]{Burada2009}
\bibinfo{author}{\bibfnamefont{P.~S.} \bibnamefont{Burada}},
  \bibinfo{author}{\bibfnamefont{P.}~\bibnamefont{Haenggi}},
  \bibinfo{author}{\bibfnamefont{F.}~\bibnamefont{Marchesoni}},
  \bibinfo{author}{\bibfnamefont{G.}~\bibnamefont{Schmid}}, \bibnamefont{and}
  \bibinfo{author}{\bibfnamefont{P.}~\bibnamefont{Talkner}},
  \bibinfo{journal}{ChemPhysChem} \textbf{\bibinfo{volume}{10}},
  \bibinfo{pages}{45} (\bibinfo{year}{2009}).

\bibitem[{\citenamefont{Scheidler et~al.}(2002)\citenamefont{Scheidler, Kob,
  and Binder}}]{Scheidler2002}
\bibinfo{author}{\bibfnamefont{P.}~\bibnamefont{Scheidler}},
  \bibinfo{author}{\bibfnamefont{W.}~\bibnamefont{Kob}}, \bibnamefont{and}
  \bibinfo{author}{\bibfnamefont{K.}~\bibnamefont{Binder}},
  \bibinfo{journal}{Europhysics Letters} \textbf{\bibinfo{volume}{59}},
  \bibinfo{pages}{701} (\bibinfo{year}{2002}).
  
  \bibitem[{\citenamefont{N\"oding et~al.}(2002)\citenamefont{N\"oding, 
  and K\"oster}}]{Noeding2012}
\bibinfo{author}{\bibfnamefont{B.}~\bibnamefont{N\"oding}}, \bibnamefont{and}
  \bibinfo{author}{\bibfnamefont{S.}~\bibnamefont{K\"oster}},
  \bibinfo{journal}{Physical Review Letters} \textbf{\bibinfo{volume}{108}},
  \bibinfo{pages}{088101} (\bibinfo{year}{2012}).

\bibitem[{\citenamefont{Bock et~al.}(2007)\citenamefont{Bock, Gubbins, and
  Schoen}}]{Bock2007}
\bibinfo{author}{\bibfnamefont{H.}~\bibnamefont{Bock}},
  \bibinfo{author}{\bibfnamefont{K.~E.} \bibnamefont{Gubbins}},
  \bibnamefont{and} \bibinfo{author}{\bibfnamefont{M.}~\bibnamefont{Schoen}},
  \bibinfo{journal}{Journal of Physical Chemistry C}
  \textbf{\bibinfo{volume}{111}}, \bibinfo{pages}{15493}
  (\bibinfo{year}{2007}).

\bibitem[{\citenamefont{Sangthong et~al.}(2008)\citenamefont{Sangthong, Probst,
  and Limtrakul}}]{Sangthong2008}
\bibinfo{author}{\bibfnamefont{W.}~\bibnamefont{Sangthong}},
  \bibinfo{author}{\bibfnamefont{M.}~\bibnamefont{Probst}}, \bibnamefont{and}
  \bibinfo{author}{\bibfnamefont{J.}~\bibnamefont{Limtrakul}},
  \bibinfo{journal}{Chemical Engineering Communications}
  \textbf{\bibinfo{volume}{195}}, \bibinfo{pages}{1486} (\bibinfo{year}{2008}).

\bibitem[{\citenamefont{Krishna}(2009)}]{Krishna2009}
\bibinfo{author}{\bibfnamefont{R.}~\bibnamefont{Krishna}},
  \bibinfo{journal}{Journal of Physical Chemistry C}
  \textbf{\bibinfo{volume}{113}}, \bibinfo{pages}{19756}
  (\bibinfo{year}{2009}).

\bibitem[{\citenamefont{Chan and Horn}(1985)}]{Chan1985}
\bibinfo{author}{\bibfnamefont{D.~Y.~C.} \bibnamefont{Chan}} \bibnamefont{and}
  \bibinfo{author}{\bibfnamefont{R.~G.} \bibnamefont{Horn}},
  \bibinfo{journal}{Journal of Chemical Physics} \textbf{\bibinfo{volume}{83}},
  \bibinfo{pages}{5311} (\bibinfo{year}{1985}).

\bibitem[{\citenamefont{Georges et~al.}(1993)\citenamefont{Georges, Millot,
  Loubet, and Tonck}}]{Georges1993}
\bibinfo{author}{\bibfnamefont{J.~M.} \bibnamefont{Georges}},
  \bibinfo{author}{\bibfnamefont{S.}~\bibnamefont{Millot}},
  \bibinfo{author}{\bibfnamefont{J.~L.} \bibnamefont{Loubet}},
  \bibnamefont{and} \bibinfo{author}{\bibfnamefont{A.}~\bibnamefont{Tonck}},
  \bibinfo{journal}{Journal of Chemical Physics} \textbf{\bibinfo{volume}{98}},
  \bibinfo{pages}{7345} (\bibinfo{year}{1993}).

\bibitem[{\citenamefont{Ruths et~al.}(1999)\citenamefont{Ruths, Ohtani,
  Greenfield, and Granick}}]{Ruths1999}
\bibinfo{author}{\bibfnamefont{M.}~\bibnamefont{Ruths}},
  \bibinfo{author}{\bibfnamefont{H.}~\bibnamefont{Ohtani}},
  \bibinfo{author}{\bibfnamefont{M.~L.} \bibnamefont{Greenfield}},
  \bibnamefont{and} \bibinfo{author}{\bibfnamefont{S.}~\bibnamefont{Granick}},
  \bibinfo{journal}{Tribology Letters} \textbf{\bibinfo{volume}{6}},
  \bibinfo{pages}{207} (\bibinfo{year}{1999}).

\bibitem[{\citenamefont{Becker and Mugele}(2003)}]{Becker2003}
\bibinfo{author}{\bibfnamefont{T.}~\bibnamefont{Becker}} \bibnamefont{and}
  \bibinfo{author}{\bibfnamefont{F.}~\bibnamefont{Mugele}},
  \bibinfo{journal}{Physical Review Letters} \textbf{\bibinfo{volume}{91}},
  \bibinfo{pages}{166104} (\bibinfo{year}{2003}).

\bibitem[{\citenamefont{Eijkel and van~den Berg}(2005)}]{Eijkel2005}
\bibinfo{author}{\bibfnamefont{J.~C.~T.} \bibnamefont{Eijkel}}
  \bibnamefont{and} \bibinfo{author}{\bibfnamefont{A.}~\bibnamefont{van~den
  Berg}}, \bibinfo{journal}{Microfluidics and Nanofluidics}
  \textbf{\bibinfo{volume}{1}}, \bibinfo{pages}{249} (\bibinfo{year}{2005}).

\bibitem[{\citenamefont{Neto et~al.}(2005)\citenamefont{Neto, Evans,
  Bonaccurso, Butt, and Craig}}]{Neto2005}
\bibinfo{author}{\bibfnamefont{C.}~\bibnamefont{Neto}},
  \bibinfo{author}{\bibfnamefont{D.~R.} \bibnamefont{Evans}},
  \bibinfo{author}{\bibfnamefont{E.}~\bibnamefont{Bonaccurso}},
  \bibinfo{author}{\bibfnamefont{H.~J.} \bibnamefont{Butt}}, \bibnamefont{and}
  \bibinfo{author}{\bibfnamefont{V.~S.~J.} \bibnamefont{Craig}},
  \bibinfo{journal}{Reports On Progress In Physics}
  \textbf{\bibinfo{volume}{68}}, \bibinfo{pages}{2859} (\bibinfo{year}{2005}).

\bibitem[{\citenamefont{Joly et~al.}(2006)\citenamefont{Joly, Ybert, and
  Bocquet}}]{Joly2006}
\bibinfo{author}{\bibfnamefont{L.}~\bibnamefont{Joly}},
  \bibinfo{author}{\bibfnamefont{C.}~\bibnamefont{Ybert}}, \bibnamefont{and}
  \bibinfo{author}{\bibfnamefont{L.}~\bibnamefont{Bocquet}},
  \bibinfo{journal}{Physical Review Letters} \textbf{\bibinfo{volume}{96}},
  \bibinfo{pages}{046101} (\bibinfo{year}{2006}).

\bibitem[{\citenamefont{Huber et~al.}(2007)\citenamefont{Huber, Gruener,
  Schaefer, Knorr, and Kityk}}]{Huber2007}
\bibinfo{author}{\bibfnamefont{P.}~\bibnamefont{Huber}},
  \bibinfo{author}{\bibfnamefont{S.}~\bibnamefont{Gruener}},
  \bibinfo{author}{\bibfnamefont{C.}~\bibnamefont{Schaefer}},
  \bibinfo{author}{\bibfnamefont{K.}~\bibnamefont{Knorr}}, \bibnamefont{and}
  \bibinfo{author}{\bibfnamefont{A.~V.} \bibnamefont{Kityk}},
  \bibinfo{journal}{European Physical Journal-special Topics}
  \textbf{\bibinfo{volume}{141}}, \bibinfo{pages}{101l} (\bibinfo{year}{2007}).

\bibitem[{\citenamefont{Kusmin et~al.}(2010{\natexlab{a}})\citenamefont{Kusmin,
  Gruener, Henschel, Holderer, Allgaier, Richter, and Huber}}]{Kusmin2010}
\bibinfo{author}{\bibfnamefont{A.}~\bibnamefont{Kusmin}},
  \bibinfo{author}{\bibfnamefont{S.}~\bibnamefont{Gruener}},
  \bibinfo{author}{\bibfnamefont{A.}~\bibnamefont{Henschel}},
  \bibinfo{author}{\bibfnamefont{O.}~\bibnamefont{Holderer}},
  \bibinfo{author}{\bibfnamefont{J.}~\bibnamefont{Allgaier}},
  \bibinfo{author}{\bibfnamefont{D.}~\bibnamefont{Richter}}, \bibnamefont{and}
  \bibinfo{author}{\bibfnamefont{P.}~\bibnamefont{Huber}},
  \bibinfo{journal}{Journal of Physical Chemistry Letters}
  \textbf{\bibinfo{volume}{1}}, \bibinfo{pages}{3116}
  (\bibinfo{year}{2010}{\natexlab{a}}).

\bibitem[{\citenamefont{Walz et~al.}(2011)\citenamefont{Walz, Gerth, Falus,
  Klimczak, Metzger, and Magerl}}]{Walz2011}
\bibinfo{author}{\bibfnamefont{M.}~\bibnamefont{Walz}},
  \bibinfo{author}{\bibfnamefont{S.}~\bibnamefont{Gerth}},
  \bibinfo{author}{\bibfnamefont{P.}~\bibnamefont{Falus}},
  \bibinfo{author}{\bibfnamefont{M.}~\bibnamefont{Klimczak}},
  \bibinfo{author}{\bibfnamefont{T.~H.} \bibnamefont{Metzger}},
  \bibnamefont{and} \bibinfo{author}{\bibfnamefont{A.}~\bibnamefont{Magerl}},
  \bibinfo{journal}{Journal of physics. Condensed matter : an Institute of
  Physics journal} \textbf{\bibinfo{volume}{23}}, \bibinfo{pages}{324102}
  (\bibinfo{year}{2011}).

\bibitem[{\citenamefont{Bocquet and Charlaix}(2010)}]{Bocquet2010}
\bibinfo{author}{\bibfnamefont{L.}~\bibnamefont{Bocquet}} \bibnamefont{and}
  \bibinfo{author}{\bibfnamefont{E.}~\bibnamefont{Charlaix}},
  \bibinfo{journal}{Chemical Society Reviews} \textbf{\bibinfo{volume}{39}},
  \bibinfo{pages}{1073} (\bibinfo{year}{2010}).

\bibitem[{\citenamefont{Gruener and Huber}(2011)}]{Gruener2011}
\bibinfo{author}{\bibfnamefont{S.}~\bibnamefont{Gruener}} \bibnamefont{and}
  \bibinfo{author}{\bibfnamefont{P.}~\bibnamefont{Huber}},
  \bibinfo{journal}{Journal of Physics-condensed Matter}
  \textbf{\bibinfo{volume}{23}}, \bibinfo{pages}{184109}
  (\bibinfo{year}{2011}).

\bibitem[{\citenamefont{Gruener and Huber}(2008)}]{Gruener2008}
\bibinfo{author}{\bibfnamefont{S.}~\bibnamefont{Gruener}} \bibnamefont{and}
  \bibinfo{author}{\bibfnamefont{P.}~\bibnamefont{Huber}},
  \bibinfo{journal}{Physical Review Letters} \textbf{\bibinfo{volume}{100}},
  \bibinfo{pages}{064502} (\bibinfo{year}{2008}).

\bibitem[{\citenamefont{Gruener et~al.}(2009)\citenamefont{Gruener, Hofmann,
  Wallacher, Kityk, and Huber}}]{Gruener2009}
\bibinfo{author}{\bibfnamefont{S.}~\bibnamefont{Gruener}},
  \bibinfo{author}{\bibfnamefont{T.}~\bibnamefont{Hofmann}},
  \bibinfo{author}{\bibfnamefont{D.}~\bibnamefont{Wallacher}},
  \bibinfo{author}{\bibfnamefont{A.~V.} \bibnamefont{Kityk}}, \bibnamefont{and}
  \bibinfo{author}{\bibfnamefont{P.}~\bibnamefont{Huber}},
  \bibinfo{journal}{Physical Review E} \textbf{\bibinfo{volume}{79}},
  \bibinfo{pages}{067301} (\bibinfo{year}{2009}).

\bibitem[{\citenamefont{Mitra et~al.}(1992)\citenamefont{Mitra, Sen, Schwartz,
  and Ledoussal}}]{Mitra1992}
\bibinfo{author}{\bibfnamefont{P.~P.} \bibnamefont{Mitra}},
  \bibinfo{author}{\bibfnamefont{P.~N.} \bibnamefont{Sen}},
  \bibinfo{author}{\bibfnamefont{L.~M.} \bibnamefont{Schwartz}},
  \bibnamefont{and}
  \bibinfo{author}{\bibfnamefont{P.}~\bibnamefont{Ledoussal}},
  \bibinfo{journal}{Physical Review Letters} \textbf{\bibinfo{volume}{68}},
  \bibinfo{pages}{3555} (\bibinfo{year}{1992}).

\bibitem[{\citenamefont{Mitra et~al.}(1993)\citenamefont{Mitra, Sen, and
  Schwartz}}]{Mitra1993}
\bibinfo{author}{\bibfnamefont{P.~P.} \bibnamefont{Mitra}},
  \bibinfo{author}{\bibfnamefont{P.~N.} \bibnamefont{Sen}}, \bibnamefont{and}
  \bibinfo{author}{\bibfnamefont{L.~M.} \bibnamefont{Schwartz}},
  \bibinfo{journal}{Physical Review B} \textbf{\bibinfo{volume}{47}},
  \bibinfo{pages}{8565} (\bibinfo{year}{1993}).

\bibitem[{\citenamefont{Chmelik et~al.}(2011)\citenamefont{Chmelik, Enke,
  Galvosas, Gobin, Jentys, Jobic, Kaerger, Krause, Kullmann, Lercher
  et~al.}}]{Chmelik2011}
\bibinfo{author}{\bibfnamefont{C.}~\bibnamefont{Chmelik}},
  \bibinfo{author}{\bibfnamefont{D.}~\bibnamefont{Enke}},
  \bibinfo{author}{\bibfnamefont{P.}~\bibnamefont{Galvosas}},
  \bibinfo{author}{\bibfnamefont{O.}~\bibnamefont{Gobin}},
  \bibinfo{author}{\bibfnamefont{A.}~\bibnamefont{Jentys}},
  \bibinfo{author}{\bibfnamefont{H.}~\bibnamefont{Jobic}},
  \bibinfo{author}{\bibfnamefont{J.}~\bibnamefont{Kaerger}},
  \bibinfo{author}{\bibfnamefont{C.~B.} \bibnamefont{Krause}},
  \bibinfo{author}{\bibfnamefont{J.}~\bibnamefont{Kullmann}},
  \bibinfo{author}{\bibfnamefont{J.}~\bibnamefont{Lercher}},
  \bibnamefont{et~al.}, \bibinfo{journal}{ChemPhysChem}
  \textbf{\bibinfo{volume}{12}}, \bibinfo{pages}{1130} (\bibinfo{year}{2011}).

\bibitem[{\citenamefont{Alder and Wainwright}(1970)}]{Alder1970}
\bibinfo{author}{\bibfnamefont{B.~J.} \bibnamefont{Alder}} \bibnamefont{and}
  \bibinfo{author}{\bibfnamefont{T.}~\bibnamefont{Wainwright}},
  \bibinfo{journal}{Physical Review A} \textbf{\bibinfo{volume}{1}},
  \bibinfo{pages}{18} (\bibinfo{year}{1970}).

\bibitem[{\citenamefont{Alder et~al.}(1970)\citenamefont{Alder, Gass, and
  Wainwright}}]{Alder1970a}
\bibinfo{author}{\bibfnamefont{B.~J.} \bibnamefont{Alder}},
  \bibinfo{author}{\bibfnamefont{D.~M.} \bibnamefont{Gass}}, \bibnamefont{and}
  \bibinfo{author}{\bibnamefont{Wainwright}}, \bibinfo{journal}{Journal of
  Chemical Physics} \textbf{\bibinfo{volume}{53}}, \bibinfo{pages}{3813}
  (\bibinfo{year}{1970}).

\bibitem[{\citenamefont{Cui}(2005)}]{Cui2005}
\bibinfo{author}{\bibfnamefont{S.~T.} \bibnamefont{Cui}},
  \bibinfo{journal}{Journal of Chemical Physics}
  \textbf{\bibinfo{volume}{123}}, \bibinfo{pages}{054706}
  (\bibinfo{year}{2005}).

\bibitem[{\citenamefont{Devi et~al.}(2010)\citenamefont{Devi, Sood, Srivastava,
  and Tankeshwar}}]{Devi2010}
\bibinfo{author}{\bibfnamefont{R.}~\bibnamefont{Devi}},
  \bibinfo{author}{\bibfnamefont{J.}~\bibnamefont{Sood}},
  \bibinfo{author}{\bibfnamefont{S.}~\bibnamefont{Srivastava}},
  \bibnamefont{and}
  \bibinfo{author}{\bibfnamefont{K.}~\bibnamefont{Tankeshwar}},
  \bibinfo{journal}{Microfluidics and Nanofluidics}
  \textbf{\bibinfo{volume}{9}}, \bibinfo{pages}{737} (\bibinfo{year}{2010}).

\bibitem[{\citenamefont{Hansen et~al.}(1998)\citenamefont{Hansen, Courivaud,
  Karlsson, Kolboe, and Stocker}}]{Hansen1998}
\bibinfo{author}{\bibfnamefont{E.~W.} \bibnamefont{Hansen}},
  \bibinfo{author}{\bibfnamefont{F.}~\bibnamefont{Courivaud}},
  \bibinfo{author}{\bibfnamefont{A.}~\bibnamefont{Karlsson}},
  \bibinfo{author}{\bibfnamefont{S.}~\bibnamefont{Kolboe}}, \bibnamefont{and}
  \bibinfo{author}{\bibfnamefont{M.}~\bibnamefont{Stocker}},
  \bibinfo{journal}{Microporous and Mesoporous Materials}
  \textbf{\bibinfo{volume}{22}}, \bibinfo{pages}{309} (\bibinfo{year}{1998}).

\bibitem[{\citenamefont{Kimmich et~al.}(1996)\citenamefont{Kimmich, Stapf,
  Maklakov, Skirda, and Khozina}}]{Kimmich1996}
\bibinfo{author}{\bibfnamefont{R.}~\bibnamefont{Kimmich}},
  \bibinfo{author}{\bibfnamefont{S.}~\bibnamefont{Stapf}},
  \bibinfo{author}{\bibfnamefont{A.~I.} \bibnamefont{Maklakov}},
  \bibinfo{author}{\bibfnamefont{V.~D.} \bibnamefont{Skirda}},
  \bibnamefont{and} \bibinfo{author}{\bibfnamefont{E.~V.}
  \bibnamefont{Khozina}}, \bibinfo{journal}{Magnetic Resonance Imaging}
  \textbf{\bibinfo{volume}{14}}, \bibinfo{pages}{793} (\bibinfo{year}{1996}).

\bibitem[{\citenamefont{Stapf et~al.}(1995)\citenamefont{Stapf, Kimmich, and
  Seitter}}]{Stapf1995}
\bibinfo{author}{\bibfnamefont{S.}~\bibnamefont{Stapf}},
  \bibinfo{author}{\bibfnamefont{R.}~\bibnamefont{Kimmich}}, \bibnamefont{and}
  \bibinfo{author}{\bibfnamefont{R.~O.} \bibnamefont{Seitter}},
  \bibinfo{journal}{Physical Review Letters} \textbf{\bibinfo{volume}{75}},
  \bibinfo{pages}{2855} (\bibinfo{year}{1995}).

\bibitem[{\citenamefont{Baumert et~al.}(2002)\citenamefont{Baumert, Asmussen,
  Gutt, and Kahn}}]{Baumert_2002_a}
\bibinfo{author}{\bibfnamefont{J.}~\bibnamefont{Baumert}},
  \bibinfo{author}{\bibfnamefont{B.}~\bibnamefont{Asmussen}},
  \bibinfo{author}{\bibfnamefont{C.}~\bibnamefont{Gutt}}, \bibnamefont{and}
  \bibinfo{author}{\bibfnamefont{R.}~\bibnamefont{Kahn}}, \bibinfo{journal}{The
  Journal of Chemical Physics} \textbf{\bibinfo{volume}{116}},
  \bibinfo{pages}{10869} (\bibinfo{year}{2002}).

\bibitem[{\citenamefont{Hair and Hertl}(1969)}]{Hair1969}
\bibinfo{author}{\bibfnamefont{M.~L.} \bibnamefont{Hair}} \bibnamefont{and}
  \bibinfo{author}{\bibfnamefont{W.}~\bibnamefont{Hertl}},
  \bibinfo{journal}{Journal of Physical Chemistry}
  \textbf{\bibinfo{volume}{73}}, \bibinfo{pages}{4269} (\bibinfo{year}{1969}).

\bibitem[{\citenamefont{Saam and Cole}(1975)}]{Saam_1975_a}
\bibinfo{author}{\bibfnamefont{W.~F.} \bibnamefont{Saam}} \bibnamefont{and}
  \bibinfo{author}{\bibfnamefont{M.~W.} \bibnamefont{Cole}},
  \bibinfo{journal}{Phys. Rev. B} \textbf{\bibinfo{volume}{11}},
  \bibinfo{pages}{1086} (\bibinfo{year}{1975}).

\bibitem[{\citenamefont{Mason}(1982)}]{Mason_1982_a}
\bibinfo{author}{\bibfnamefont{G.}~\bibnamefont{Mason}},
  \bibinfo{journal}{Journal of Colloid and Interface Science}
  \textbf{\bibinfo{volume}{88}}, \bibinfo{pages}{36 } (\bibinfo{year}{1982}),
  ISSN \bibinfo{issn}{0021-9797}.

\bibitem[{\citenamefont{Mason}(1983)}]{Mason_1983_a}
\bibinfo{author}{\bibfnamefont{G.}~\bibnamefont{Mason}},
  \bibinfo{journal}{Proceedings of the Royal Society of London. A. Mathematical
  and Physical Sciences} \textbf{\bibinfo{volume}{390}}, \bibinfo{pages}{47}
  (\bibinfo{year}{1983}).

\bibitem[{\citenamefont{Kornev et~al.}(2002)\citenamefont{Kornev, Shingareve,
  and Neimark}}]{Kornev_2002_a}
\bibinfo{author}{\bibfnamefont{K.~G.} \bibnamefont{Kornev}},
  \bibinfo{author}{\bibfnamefont{I.~K.} \bibnamefont{Shingareve}},
  \bibnamefont{and} \bibinfo{author}{\bibfnamefont{A.~V.}
  \bibnamefont{Neimark}}, \bibinfo{journal}{ADvances in Volloid and Interface
  Science} \textbf{\bibinfo{volume}{96}}, \bibinfo{pages}{143}
  (\bibinfo{year}{2002}).

\bibitem[{\citenamefont{Knorr et~al.}(2008)\citenamefont{Knorr, Huber, and
  Wallacher}}]{Knorr_2008_a}
\bibinfo{author}{\bibfnamefont{K.}~\bibnamefont{Knorr}},
  \bibinfo{author}{\bibfnamefont{P.}~\bibnamefont{Huber}}, \bibnamefont{and}
  \bibinfo{author}{\bibfnamefont{D.}~\bibnamefont{Wallacher}},
  \bibinfo{journal}{Zeitschfrift f\"ur Physikalische Chemie}
  \textbf{\bibinfo{volume}{222}}, \bibinfo{pages}{257} (\bibinfo{year}{2008}).

\bibitem[{\citenamefont{Henschel et~al.}(2009)\citenamefont{Henschel, Kumar,
  Hofmann, Knorr, and Huber}}]{Henschel_2009_a}
\bibinfo{author}{\bibfnamefont{A.}~\bibnamefont{Henschel}},
  \bibinfo{author}{\bibfnamefont{P.}~\bibnamefont{Kumar}},
  \bibinfo{author}{\bibfnamefont{T.}~\bibnamefont{Hofmann}},
  \bibinfo{author}{\bibfnamefont{K.}~\bibnamefont{Knorr}}, \bibnamefont{and}
  \bibinfo{author}{\bibfnamefont{P.}~\bibnamefont{Huber}},
  \bibinfo{journal}{Physical Review E} \textbf{\bibinfo{volume}{79}},
  \bibinfo{pages}{032601} (\bibinfo{year}{2009}).

\bibitem[{\citenamefont{Furrer et~al.}(2009)\citenamefont{Furrer, Mesot, and
  Str\"assle}}]{Furrer_2009_a}
\bibinfo{author}{\bibfnamefont{A.}~\bibnamefont{Furrer}},
  \bibinfo{author}{\bibfnamefont{J.}~\bibnamefont{Mesot}}, \bibnamefont{and}
  \bibinfo{author}{\bibfnamefont{T.}~\bibnamefont{Str\"assle}},
  \emph{\bibinfo{title}{Neutron Scattering in Condensed Matter Physics}}
  (\bibinfo{publisher}{World Scientific}, \bibinfo{year}{2009}).

\bibitem[{\citenamefont{Chudley and Elliott}(1961)}]{Chudley_1961_a}
\bibinfo{author}{\bibfnamefont{C.~T.} \bibnamefont{Chudley}} \bibnamefont{and}
  \bibinfo{author}{\bibfnamefont{R.~J.} \bibnamefont{Elliott}},
  \bibinfo{journal}{Proceedings of the Physical Society}
  \textbf{\bibinfo{volume}{77}}, \bibinfo{pages}{353} (\bibinfo{year}{1961}).

\bibitem[{\citenamefont{Smuda et~al.}(2008)\citenamefont{Smuda, Busch,
  Gemmecker, and Unruh}}]{Smuda_2008_a}
\bibinfo{author}{\bibfnamefont{C.}~\bibnamefont{Smuda}},
  \bibinfo{author}{\bibfnamefont{S.}~\bibnamefont{Busch}},
  \bibinfo{author}{\bibfnamefont{G.}~\bibnamefont{Gemmecker}},
  \bibnamefont{and} \bibinfo{author}{\bibfnamefont{T.}~\bibnamefont{Unruh}},
  \bibinfo{journal}{The Journal of Chemical Physics}
  \textbf{\bibinfo{volume}{129}}, \bibinfo{pages}{014513}
  (\bibinfo{year}{2008}).

\bibitem[{\citenamefont{Bee}(1992)}]{Bee_1992_a}
\bibinfo{author}{\bibfnamefont{M.}~\bibnamefont{Bee}},
  \bibinfo{journal}{Physica B} \textbf{\bibinfo{volume}{182}},
  \bibinfo{pages}{323} (\bibinfo{year}{1992}).

\bibitem[{\citenamefont{Douglass and McCall}(1958)}]{Douglass_1958_a}
\bibinfo{author}{\bibfnamefont{D.~C.} \bibnamefont{Douglass}} \bibnamefont{and}
  \bibinfo{author}{\bibfnamefont{D.~W.} \bibnamefont{McCall}},
  \bibinfo{journal}{The Journal of Physical Chemistry}
  \textbf{\bibinfo{volume}{62}}, \bibinfo{pages}{1102} (\bibinfo{year}{1958}),
  ISSN \bibinfo{issn}{0022-3654}.

\bibitem[{\citenamefont{McCall et~al.}(1959)\citenamefont{McCall, Douglass, and
  Anderson}}]{McCall_1959_a}
\bibinfo{author}{\bibfnamefont{D.~W.} \bibnamefont{McCall}},
  \bibinfo{author}{\bibfnamefont{D.~C.} \bibnamefont{Douglass}},
  \bibnamefont{and} \bibinfo{author}{\bibfnamefont{E.~W.}
  \bibnamefont{Anderson}}, \bibinfo{journal}{The Physics of Fluids}
  \textbf{\bibinfo{volume}{2}}, \bibinfo{pages}{87} (\bibinfo{year}{1959}).

\bibitem[{\citenamefont{Dianoux et~al.}(1975)\citenamefont{Dianoux, Volino, and
  Hervet}}]{Dianoux_1975_a}
\bibinfo{author}{\bibfnamefont{A.~J.} \bibnamefont{Dianoux}},
  \bibinfo{author}{\bibfnamefont{F.}~\bibnamefont{Volino}}, \bibnamefont{and}
  \bibinfo{author}{\bibfnamefont{H.}~\bibnamefont{Hervet}},
  \bibinfo{journal}{Molecular Physics: An International Journal at the
  Interface Between Chemistry and Physics} \textbf{\bibinfo{volume}{30}},
  \bibinfo{pages}{1181} (\bibinfo{year}{1975}), ISSN \bibinfo{issn}{0026-8976}.

\bibitem[{\citenamefont{Zorn et~al.}(2002)\citenamefont{Zorn, Frick, and
  Fetters}}]{Zorn_2002_a}
\bibinfo{author}{\bibfnamefont{R.}~\bibnamefont{Zorn}},
  \bibinfo{author}{\bibfnamefont{B.}~\bibnamefont{Frick}}, \bibnamefont{and}
  \bibinfo{author}{\bibfnamefont{L.~J.} \bibnamefont{Fetters}},
  \bibinfo{journal}{J. Chem. Phys.} \textbf{\bibinfo{volume}{116}},
  \bibinfo{pages}{845} (\bibinfo{year}{2002}).

\bibitem[{\citenamefont{Volkmann et~al.}(2002)\citenamefont{Volkmann, Pino,
  Altamirano, Taub, and Hansen}}]{Volkmann2002}
\bibinfo{author}{\bibfnamefont{U.~G.} \bibnamefont{Volkmann}},
  \bibinfo{author}{\bibfnamefont{M.}~\bibnamefont{Pino}},
  \bibinfo{author}{\bibfnamefont{L.~A.} \bibnamefont{Altamirano}},
  \bibinfo{author}{\bibfnamefont{H.}~\bibnamefont{Taub}}, \bibnamefont{and}
  \bibinfo{author}{\bibfnamefont{F.~Y.} \bibnamefont{Hansen}},
  \bibinfo{journal}{Journal of Chemical Physics}
  \textbf{\bibinfo{volume}{116}}, \bibinfo{pages}{2107} (\bibinfo{year}{2002}).

\bibitem[{\citenamefont{Schranz et~al.}(2007)\citenamefont{Schranz, Puica,
  Koppensteiner, Kabelka, and Kityk}}]{Schranz2007}
\bibinfo{author}{\bibfnamefont{W.}~\bibnamefont{Schranz}},
  \bibinfo{author}{\bibfnamefont{M.~R.} \bibnamefont{Puica}},
  \bibinfo{author}{\bibfnamefont{J.}~\bibnamefont{Koppensteiner}},
  \bibinfo{author}{\bibfnamefont{H.}~\bibnamefont{Kabelka}}, \bibnamefont{and}
  \bibinfo{author}{\bibfnamefont{A.~V.} \bibnamefont{Kityk}},
  \bibinfo{journal}{Epl} \textbf{\bibinfo{volume}{79}}, \bibinfo{pages}{36003}
  (\bibinfo{year}{2007}).

\bibitem[{\citenamefont{Courivaud et~al.}(2000)\citenamefont{Courivaud, Hansen,
  Karlsson, Kolboe, and Stocker}}]{Courivaud2000}
\bibinfo{author}{\bibfnamefont{F.}~\bibnamefont{Courivaud}},
  \bibinfo{author}{\bibfnamefont{E.~W.} \bibnamefont{Hansen}},
  \bibinfo{author}{\bibfnamefont{A.}~\bibnamefont{Karlsson}},
  \bibinfo{author}{\bibfnamefont{S.}~\bibnamefont{Kolboe}}, \bibnamefont{and}
  \bibinfo{author}{\bibfnamefont{M.}~\bibnamefont{Stocker}},
  \bibinfo{journal}{Microporous and Mesoporous Materials}
  \textbf{\bibinfo{volume}{35-6}}, \bibinfo{pages}{327} (\bibinfo{year}{2000}).

\bibitem[{\citenamefont{Kityk et~al.}(2009)\citenamefont{Kityk, Knorr, and
  Huber}}]{Kityk2009}
\bibinfo{author}{\bibfnamefont{A.~V.} \bibnamefont{Kityk}},
  \bibinfo{author}{\bibfnamefont{K.}~\bibnamefont{Knorr}}, \bibnamefont{and}
  \bibinfo{author}{\bibfnamefont{P.}~\bibnamefont{Huber}},
  \bibinfo{journal}{Physical Review B} \textbf{\bibinfo{volume}{80}},
  \bibinfo{pages}{035421} (\bibinfo{year}{2009}).

\bibitem[{\citenamefont{Wolff et~al.}(2010)\citenamefont{Wolff, Knorr, Huber,
  and Kityk}}]{Wolff2010}
\bibinfo{author}{\bibfnamefont{M.}~\bibnamefont{Wolff}},
  \bibinfo{author}{\bibfnamefont{K.}~\bibnamefont{Knorr}},
  \bibinfo{author}{\bibfnamefont{P.}~\bibnamefont{Huber}}, \bibnamefont{and}
  \bibinfo{author}{\bibfnamefont{A.~V.} \bibnamefont{Kityk}},
  \bibinfo{journal}{Physical Review B} \textbf{\bibinfo{volume}{82}},
  \bibinfo{pages}{235404} (\bibinfo{year}{2010}).

\bibitem[{\citenamefont{Kusmin et~al.}(2010{\natexlab{b}})\citenamefont{Kusmin,
  Gruener, Henschel, de~Souza, Allgaier, Richter, and Huber}}]{Kusmin2010a}
\bibinfo{author}{\bibfnamefont{A.}~\bibnamefont{Kusmin}},
  \bibinfo{author}{\bibfnamefont{S.}~\bibnamefont{Gruener}},
  \bibinfo{author}{\bibfnamefont{A.}~\bibnamefont{Henschel}},
  \bibinfo{author}{\bibfnamefont{N.}~\bibnamefont{de~Souza}},
  \bibinfo{author}{\bibfnamefont{J.}~\bibnamefont{Allgaier}},
  \bibinfo{author}{\bibfnamefont{D.}~\bibnamefont{Richter}}, \bibnamefont{and}
  \bibinfo{author}{\bibfnamefont{P.}~\bibnamefont{Huber}},
  \bibinfo{journal}{Macromolecules} \textbf{\bibinfo{volume}{43}},
  \bibinfo{pages}{8162} (\bibinfo{year}{2010}{\natexlab{b}}).

\end{thebibliography}
\end{document}